\documentstyle[prb,aps,floats,graphicx]{revtex}
\def\et{{\it et al.}}
	
\newcommand{\be}{\begin{equation}}
\newcommand{\ee}{\end{equation}}
\newcommand{\ber}{\begin{eqnarray}}
\newcommand{\eer}{\end{eqnarray}}

\begin{document}

\pagestyle{plain}

\twocolumn[\hsize\textwidth\columnwidth\hsize\csname
@twocolumnfalse\endcsname






\title{Smectic ordering in liquid crystal - aerosil dispersions I. X-ray scattering}

\author{R. L. Leheny$^{1}$, S. Park$^{2}$\thanks{Present Address: 
NCNR, NIST, Gaithersburg, MD.}, R. J. Birgeneau$^{2,3}$, 
J.-L. Gallani$^{2}$\thanks{Permanent Address: 
IPCMS-GMO, Strasbourg, France.}, C. W. Garland$^{2}$, 
and G. S. Iannacchione$^{4}$}

\address{$^{1}$Department of Physics and Astronomy, Johns Hopkins
University,
Baltimore, MD 21218, USA
\\
$^{2}$Center for Materials Science and Engineering, 
Massachusetts Institute of Technology,
Cambridge, MA 02139, USA
\\
$^{3}$Department of Physics, University of Toronto, Toronto, Canada 
M5S 1A7
\\
$^{4}$Department of Physics, Worcester Polytechnic Institute,
Worcester, MA 01609, USA
}

\date{\today}
\maketitle

\begin{abstract}
Comprehensive x-ray scattering studies have characterized the smectic 
ordering of octylcyanobiphenyl (8CB) confined in the hydrogen-bonded silica 
gels formed by aerosil dispersions.  For all densities of aerosil and all
measurement temperatures, the correlations remain short range, 
demonstrating that the disorder imposed by the gels destroys 
the nematic (N) to smectic-A (SmA) transition.  The smectic correlation function contains two 
distinct contributions.  The 
first has a form identical to that describing the critical 
thermal fluctuations in pure 8CB near the N-SmA transition, and this 
term displays a 
temperature dependence at high temperatures similar to that of 
the pure liquid crystal.  The second term, which is negligible at 
high temperatures but dominates at low 
temperatures, has a 
shape given by the thermal term squared and describes the 
static fluctuations due to random fields induced by confinement in the 
gel.  The correlation lengths appearing in the thermal and disorder 
terms are the same and show strong variation with gel density at low 
temperatures.
The temperature dependence of the amplitude of the static 
fluctuations further suggests that nematic 
susceptibility become suppressed with increasing quenched disorder.
The results overall are well described by a mapping of the liquid
crystal-aerosil system into a three dimensional XY model in a random field
with disorder strength varying linearly with the aerosil density.

\end{abstract}
\


\narrowtext
\vskip0pc]
\newpage

\section {Introduction}

Liquid crystals have long served as important model systems in 
statistical mechanics.  For example, experiments on phase transitions
in liquid crystals have provided many of the most detailed tests of the modern theories of 
critical phenomena.  Recent studies in liquid crystals have investigated the effects that 
quenched disorder produces in phase 
behavior and mesophase ordering, and a
fruitful strategy in experiments for introducing quenched disorder 
has been through confinement in random 
porous media.  The fragile nature of the mesomorphic phases 
and the importance of surface interactions make such confinement 
particularly well suited for liquid crystalline systems.  For example, 
a series of 
experiments~\cite{lei,bellini,rappaport,finotello-gel,bellini-aerogel-light}
has characterized the effects 
on the nematic to isotropic and nematic to smectic-A transitions of liquid 
crystals confined in aerogels -- highly-porous, chemically bonded silica gels.  
The primary conclusion of this work has been that 
the transition behavior becomes severely smeared and that long-range nematic and smectic ordering 
is suppressed by the aerogel.  In particular, these studies have verified 
theoretical expectations about the fragility of the smectic 
phase to disorder~\cite{clark-science}.

In order to access a weaker regime of disorder than is possible with aerogel or other 
rigid porous media, several recent studies have focused on liquid crystals confined
\newline 
\_\_\_\_\_\_\_\_\_\_\_\_\_\_\_\_
\newline
$^{*}$ Present Address: NCNR, NIST, Gaithersburg, MD.
\newline
$^{\dagger}$ Permanent Address: IPCMS-GMO, Strasbourg, France.
\newline

\noindent
 within aerosil 
gels -- weak, 
thixotropic gels comprised of nanometer scale silica particles.  
The disorder imposed by the aerosil gels can be made less severe than 
in the case of
aerogels both because lower volume fractions are 
possible with aerosils and because the compliance of the aerosil gels 
leads to partial annealing of the disorder.  This 
work on liquid crystals in aerosil dispersions, including
calorimetry~\cite{germano}, NMR~\cite{finotello-sil}, dielectric 
susceptibility~\cite{theon}, and static light 
scattering~\cite{bellini-sil-PRE}, 
has revealed important differences between such systems and liquid 
crystals in aerogels, which can be 
ascribed to the weaker nature of the aerosil disorder.  In particular, 
an x-ray scattering investigation into the 
nematic to smectic-A transition in a prototypical thermotropic liquid 
crystal, octylcyanobiphenyl (8CB), in the presence of dispersed
aerosil gels has shown
that the aerosil gel can be 
quantitatively understood as introducing weak-to-intermediate 
strength random fields that compete with smectic ordering 
in the liquid crystal~\cite{park}.   

Random field sytems have been 
fruitful models for exploring effects of quenched 
disorder, and experimentally random field Ising magnets have 
been at the center of this work~\cite{birgeneau}.  
The nematic to smectic transition breaks a 3D-XY symmetry, thus 8CB with dispersed aerosils 
provides an opportunity to study experimentally a random field 
system that breaks 
a continuous 
symmetry~\cite{imry-ma,aharony&pytte,aizenman,fisher1,leo,ward}.  
In this paper we provide a comprehensive 
picture of the smectic correlations in 8CB confined in aerosil gels, as 
determined by the x-ray scattering.  The analysis we apply adopts 
a random-field picture, and consistent 
with theoretical predictions for a 3D-XY system with random fields, 
we find that the smectic phase is 
destroyed by the disorder and is replaced by the growth of short-range smectic 
correlations.   A noteworthy feature of these x-ray results and 
corresponding calorimetric studies~\cite{germano} is the lack of any measurable
hysteresis or time dependent effects that would indicate 
out-of-equilibrium behavior.  In 
the random field Ising magnet slow dynamics and 
metastability~\cite{fisher,feng}
severely complicate efforts to understand the underlying equilibrium 
behavior.  The absence of such problems for the smectic in aerosil, 
presumably a consequence of the system's continuous symmetry, thus 
provides a unique perspective on equilibrium behavior in a random 
field system.
The companion paper that follows this 
paper~\cite{paperII}, hereafter called Paper II, compares the x-ray and
calorimetry results to reveal 
scaling behavior, including finite-size scaling effects, in 
the smectic ordering due to the presence of the aerosil gel.

Recent theoretical work has emphasized the strong effect of confinement in random media
on liquid crystal phases~\cite{leo,ward,feldman,palffymuhoray,maritan}.  A detailed study by 
Radzihovsky and Toner~\cite{leo}, which models the confinement as 
introducing random fields,
has predicted the destruction of the smectic-A phase by arbitrarily weak 
quenched disorder, consistent with experiment,
and has introduced the possibility that a topologically ordered
``smectic Bragg glass'' phase may appear at low temperatures.  
The smectic Bragg glass is a possible manifestation of the anomalous 
elasticity that the theory predicts for the smectic in the presence of
disorder.  Distinguishing features of this anomalous elasticity and smectic Bragg 
glass phase are the smectic correlations in the system.  
Thus, x-ray scattering, which can directly probe the 
smectic correlation function, is an ideal probe for testing some of 
these ideas, and recent analysis of x-ray studies on 8CB in aerogels has 
been reported to show agreement 
with many of these predictions~\cite{clark-science}.  The weaker nature of  
disorder from aerosil gels should place 8CB + aerosil samples more firmly in 
the regime addressed by the theory.  However, as shown in the present 
work, the agreement between the observed smectic correlations in 8CB in aerosil 
gels and more detailed predictions of the theory is limited.

Section II of this paper describes the sample preparation and the 
details of the x-ray scattering measurements on 8CB 
in aerosil gels, and Sec.~III outlines the results of the x-ray lineshape 
analysis, which yields insight into both thermal and static smectic 
fluctuations and the temperature dependence of correlation lengths and 
smectic susceptibilities.  Section IV provides a discussion 
of the pseudo-critical structural behavior of 8CB in weak aerosil gels.  The latter 
aspects are extended by the scaling behavior and comparisons with 
calorimetry covered in Paper II.

\vspace{1 cm}
\section {Experimental Procedures}

\subsection {Sample preparation and characteristics}

The 8CB used in this study came from a 
single synthetic batch obtained from Aldrich Corp. The liquid crystal had a quoted purity of 
99\% and was used without further 
purification.  In the absence of disorder, pure 8CB undergoes an isotropic (I) to 
nematic (N) transition at 
$T_{NI}^{0}= 313.98$ K and a nematic to smectic-A (SmA) transition at 
$T_{NA}^{0}= 306.97$ K~\cite{germano}.  Below 290 K, 8CB forms a 
three-dimensional ordered crystal phase.  The aerosil, obtained from Degussa 
Corp., consists of 70-{\AA} diameter SiO$_{2}$ spheres, and the
type 300 aerosil used in this study is 
strongly hydrophilic. 

The aerosil gels were formed directly in the liquid crystal following established 
procedures~\cite{germano}.  Appropriate quantities of degassed 8CB and 
dried aerosil 
powder were mixed with high 
purity acetone, and the suspension was sonicated for several hours to achieve a uniform 
dispersion.  The suspension was then gently heated to 315 K 
to evaporate the acetone slowly.  After no signs of acetone remained, 
the samples were placed under vacuum at 10$^{-2}$ Torr for 12 hours at 
320 K to 
remove any trace amounts of 
solvent or absorbed water vapor.  For aerosil densities above a 
gelation threshold of approximately 1\% silica by volume, the resulting 
material was a highly uniform soft solid that maintained its shape 
when heated above $T_{NI}^{0}$.  Small-angle 
x-ray scattering has revealed that aerosil gels formed under this 
procedure bear a strong 
resemblance to rigid aerogels~\cite{lei,germano}.  An extended discussion
of the gel structure formed by aerosils is given in Paper II.  The samples in this study ranged 
in aerosil densities from $\rho_{S} = 0.025$ g sil/cm$^{3}$ 8CB, 
which is just above the gelation threshold, to 
$\rho_{S} = 0.341$ g sil/cm$^{3}$ 8CB.  
Several efforts to prepare samples below the threshold density produced 
macroscopically inhomogeneous materials that were unsuitable for study. 

\subsection {X-ray scattering}

The x-ray scattering studies were conducted on the X20A and X20C 
beamlines of the National Synchrotron Light Source at Brookhaven 
National Laboratory using 8 keV x-rays.  The samples were placed 
in aluminum holders with epoxy-sealed Kapton windows designed to 
maintain a sample thickness of approximately 1 mm, closely matching the 
attenuation length for the 8 keV radiation through the material.  The 
holders were mounted in a brass block which in turn was 
positioned in a beryllium can containing dry nitrogen gas.  
A thermoelectric cooler mounted on top of the beryllium can allowed us 
to cool the sample below room temperature when desired, and a home-built P-I 
temperature controller maintained a set temperature for the brass 
block.  Preliminary x-ray scattering 
measurements
showed that the crystallization of 8CB was suppressed 
approximately 7 K by 
confinement in aerosil gels.  Because such crystallization irreversibly 
damages the aerosil gels, we limited our studies to temperatures 
safely above this point.  The measurement temperatures ranged from 287 K, approximately 20 K below 
the N-SmA transition in pure 8CB, to 318 K,
approximately 4 K above the N-I transition.  The temperature stability during a 
measurement was better than $\pm$ 0.001 K.

The scattering intensity was measured in transmission in a vertical 
scattering geometry.  The scattered beam was reflected from a 
single-crystal germanium analyzer to achieve high wavevector resolution 
and was measured with a point scintillation detector.
Because the scattering intensity from the liquid 
crystal-aerosil gel composites had the azimuthal 
symmetry of a powder pattern, two sets of slits between the sample and 
the analyzer were used to define tightly the horizontal acceptance 
and thus provide an undistorted measurement of the scattering 
intensity as a function of wavevector transfer, $I(q)$~\cite{aperature}. 

\begin{figure}[top]
    \centering\includegraphics[scale=0.8]{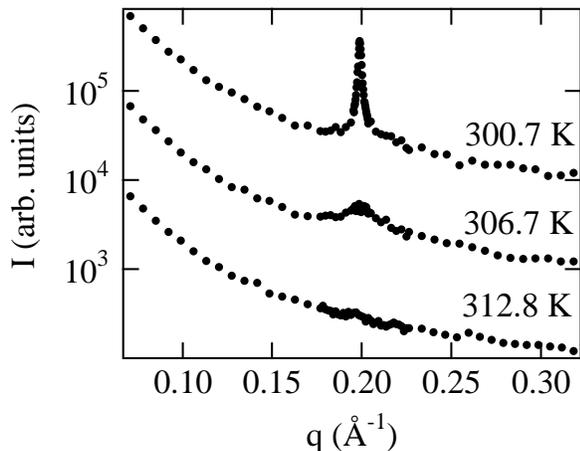}
\caption{Scattering intensity for 8CB in an aerosil gel with $\rho_{S}=0.161$ g/cm$^{3}$ at three 
temperatures spanning the N-SmA transition of the pure
system, $T_{NA}^{0}=306.97$ K.  The curves for 300.7 K and 306.7 K 
have been shifted by factors of 100 and 10, respectively, for clarity.
The strongly temperature dependent 
peak near $q = 0.2$ \AA$^{-1}$ corresponds to scattering from smectic 
fluctuations.}
\label{fig1}
\end{figure}

Preliminary measurements of the scattering from 8CB + aerosil samples
revealed damage to the material from prolonged exposure to the x-ray 
radiation.  This damage was reflected in irreproducibility in the 
observed $I(q)$ and a progressive shift in
the onset of the short-ranged smectic order described in Section 
IIID.  To combat this problem a protocol was developed which made 
these effects immeasurably small.  Immediately after the 
sample was loaded into the 
scattering cell and before any x-ray measurements were made, the material was heated 
to approximately 318 K, several degrees above $T_{NI}^{0}$.  The material was then 
cooled through a series of temperatures at which measurements 
of the scattering intensity, $I(q)$, were made.  A process of trial-and-error 
during preliminary studies revealed the appropriate exposure 
times for the final measurements to optimize statistics while avoiding 
measurable damage.  After the series of measurements ending at the lowest temperature 
was complete, the material was reheated and measurements were repeated at several 
temperatures to assure that no measurable changes in 
the scattering had occurred.  Finally, the material was heated back to 
318 K, and a measurement with a long counting time was 
conducted to determine the background scattering from the aerosil gel structure.  

\section {Results}

\subsection{Lineshape analysis}

Figure 1 displays representative results for $I(q)$ from a sample with 
$\rho_{S}=0.161$ g/cm$^{3}$
at three temperatures spanning the N-SmA 
transition temperature in pure 8CB, $T_{NA}^{0}=306.97$ K.  The scattering lineshape contains 
two salient features: a broad, sloping background that is 
approximately temperature independent and a strongly temperature-dependent 
peak that develops on cooling near 0.2 \AA$^{-1}$.  The background 
consists primarily of scattering from the aerosil gel structure.  The temperature-dependent peak 
results from smectic fluctuations in the 8CB.  The shape of this peak 
is a direct measure of the smectic correlation function for the 8CB 
confined in the aerosil gel.

\begin{figure}
    \centering\includegraphics[scale=0.8]{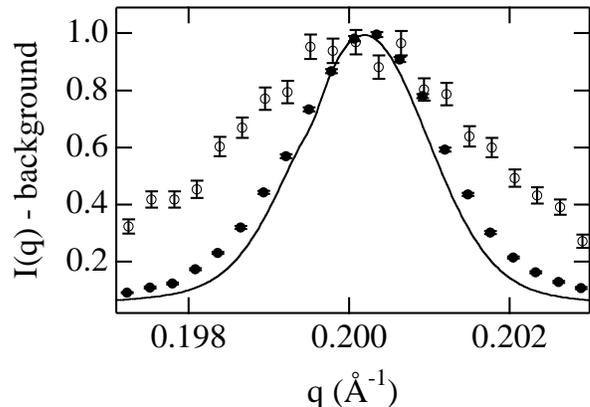}
\caption{Comparison of the smectic Bragg peak measured for 8CB in aerosil 
gels
with that measured for pure 8CB.  All peak heights have been 
normalized to 1.  The solid line is for pure 8CB below 
the N-SmA transition temperature.  The solid circles correspond to $\rho_{S}=0.025$ g/cm$^{3}$ and 
the open circles to $\rho_{S}=0.220$ g/cm$^{3}$.  The width of the peak for pure 8CB 
is essentially identical to that set by the experimental resolution, 
as determined from the direct x-ray beam profile.  The wider peaks 
measured for 8CB in aerosil gels indicate that the smectic correlations 
remain short-ranged to low temperatures (approximately $T_{NA}^{0}-19 
K$ shown here) due to the disorder imposed by the gels.}
\label{fig2}
\end{figure}

For the full range of aerosil
densities and to the lowest measurement temperatures, the observed smectic 
peak is significantly 
broader than the resolution limit set by the x-ray optics, as 
illustrated in Figure 2.  Measurements of pure 8CB 
below $T_{NA}^{0}$, shown by the solid line in the figure, revealed a 
lineshape whose width is virtually 
indistinguishable from that of the resolution, which we determined from 
the shape of the profile of the direct x-ray beam.  (The subtle increase 
in width in pure 8CB due to Landau-Pieirls instability is not 
resolved.) The solid circles in Fig.~2 are the lineshape measured 
for an 8CB + aerosil sample with $\rho_{S}=0.025$, the lowest 
aerosil density, at 287.8 K 
($T_{NA}^{0}-19.2$ K).  The larger width of the aerosil peak with respect to the resolution
reflects the finite extent of the smectic correlations
in the presence of the aerosil gel, and the consequent destruction of 
the transition to a quasi-long-range ordered smectic 
state by the quenched disorder.  The length scale 
for this short-range order depends strongly on $\rho_{S}$, as 
revealed by the contrast between the low-temperature width for the
$\rho_{S}=0.025$ sample and that for a $\rho_{S}=0.220$ sample, shown 
by the open circles in Fig.~2.

To characterize these short-range correlations, we model the structure 
factor for the smectic ordering with the form

\begin{eqnarray}
    S({\bf q}) = 
    \frac{\sigma_{1}}{1+(q_{\|}-q_{0})^{2}\xi_{\|}^{2}
    +q_{\bot}^{2}\xi_{\bot}^{2}+cq_{\bot}^{4}\xi_{\bot}^{4}} \nonumber\\ 
 + \frac{a_{2}(\xi_{\|}\xi_{\bot}^{2})}{(1+(q_{\|}-q_{0})^{2}\xi_{\|}^{2}+q_{\bot}^{2}\xi_{\bot}^{2}+
    cq_{\bot}^{4}\xi_{\bot}^{4})^{2}}
\end{eqnarray} 
where the wave vector $q_{\|}$ is along the smectic layer normal and $q_{\bot}$ 
is perpendicular to it.  The first term in $S(q)$, an anisotropic Lorentzian with quartic corrections, 
has the same form as the structure factor in pure nematic 8CB 
and represents the smectic susceptibility, with $\sigma_{1}$ equal to 
the magnitude of the susceptibility at the ordering wavevector, $q_{0}$.  This 
term characterizes the critical thermal fluctuations on approaching 
the N-SmA transition~\cite{ocko}.  
The second term, whose shape is given by the susceptibility squared, is designed to 
account for the static fluctuations induced by the quenched disorder.  
This second term is motivated by studies on random field 
systems.  Such an expression, known as the 
disconnected susceptibility, has been shown to describe accurately the short-range correlations 
induced by the static fluctuations in random field Ising magnets~\cite{RFIMexpt} and has been justified 
theoretically~\cite{grinstein,joanny,aharony} for these systems.  A very 
similar expression has also been derived specifically for smectic 
liquid crystals in the presence of random fields~\cite{ward}.  In the limit of 
long-range order, the second term evolves into a Bragg peak, 
$a_{2}\delta(q_{\|}-q_{0})\delta(q_{\bot})$~\cite{quasi}.  In applying 
Eq.~(1) to the smectic correlations of 8CB in aerosil gels, 
$\xi_{\bot}$ and $c$ are treated as functions of 
$\xi_{\|}$, with $\xi_{\bot}(\xi_{\|})$ and $c(\xi_{\|})$ 
set by their relations in pure 8CB, which are known to high 
precision~\cite{ocko}.

The total measured scattering intensity is fit to the powder average 
of $S({\bf q})$ from Eq.~(1), convolved with the resolution, plus a term to account for the background 
scattering from the aerosil gel shown in Fig.~1:
\begin{equation}
    I(q)=\int dq' \int d\Omega S({\bf q'})Res(q-q')+A(T)B(q)
\end{equation}
The powder average, represented by the integral over solid angles, can 
be solved analytically for the $S({\bf q})$ form given by 
Eq.~(1).  The details of this calculation are provided in the 
Appendix.  The convolution 
with the resolution function, $Res(q)$, was performed numerically.  The 
shape of the background, $B(q)$,
was taken from measurements at a high temperature, where scattering from the aerosil 
gel dominates, and this quantity is multiplied by a temperature 
dependent parameter, $A(T)$, which accounts for variations in the 
scattering contrast between the silica and 8CB due to differences in 
their
thermal expansion.   We found that $A(T)$ typically differed from unity by less than 
10\% 
over the full temperature range of the measurements.  

\begin{figure}
    \centering\includegraphics[scale=0.8]{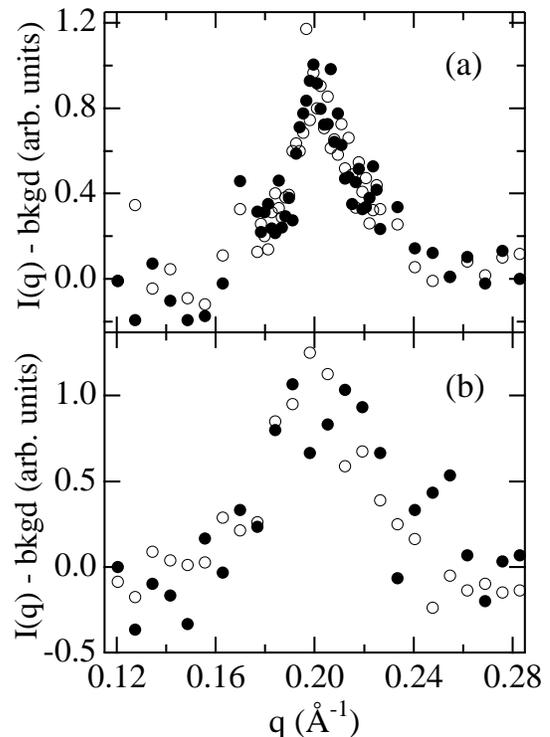}
\caption{Comparison of the smectic peaks measured for 8CB in aerosil 
gels and those measured for pure 8CB above $T_{NA}^{0}=306.97$ K: (a)  
$\rho_{S}=0.078$ g/cm$^{3}$ (solid circles) and pure 8CB (open circles) 
each at 308.3 K; and (b) $\rho_{S}=0.031$ g/cm$^{3}$ 
(solid circles) at 310.8 K
and pure 8CB (open circles) at 310.5 K.}
\label{fig3}
\end{figure}

At temperatures near and above $T_{NA}^{0}$, the measured lineshapes 
in the aerosil samples are described very well by the thermal fluctuation term alone 
(i.e., fitted $a_{2}$ are very small, and one can set $a_{2} = 0$ with 
almost no effect on the other parameters).  Thus, for $T > T_{NA}^{0}$, the smectic 
correlations in aerosil samples are very similar to the critical fluctuations of the pure liquid 
crystal.  This strong similarity is demonstrated in Fig.~3 which 
compares the smectic peak for two values of $\rho_{S}$ with that 
measured for pure 8CB at temperatures 
several degrees above $T_{NA}^{0}$.  

We note that the lineshapes of 
the smectic peaks in Figs.~1-3 are nearly symmetric in $q$ about the peak 
position.  Such highly symmetric shapes are, in general, not expected 
from powder averaging Eq.~(1).  We interpret this symmetry as a consequence of the relative 
values of $\xi_{\bot}$, $\xi_{\|}$, and $c$, particular to 8CB, which conspire to produce
such a shape.  Indeed, this symmetry highlights the importance of including the quartic term ($c>0$) in 
Eq.~(1) to account correctly 
for the measured lineshapes.  To illustrate this point, Fig.~4 shows the 
measured scattering peak from smectic fluctuations in 
pentylphenylthiol-octyloxybenzoate ($\overline{8}$S5) confined in an aerosil gel with 
$\rho_{S}=0.030$ at 2.56 K {\it above} the N-SmA transition 
temperature of pure $\overline{8}$S5.  For a given $\xi_{\|}$, the quartic term $c$
is considerably smaller in $\overline{8}$S5 than in 8CB.  This difference leads to 
a much more asymmetric shape to the powder average of the thermal term 
in Eq.~(1).  The solid line in Fig.~4 is a fit with Eqs.~(1) and (2) with 
$a_{2}=0$~\cite{8s5fits}, so that only the thermal term is 
included in describing the smectic fluctuations above $T_{NA}$.

In measurements on pure 8CB above $T_{NI}^{0}$, we observe a very broad liquid structure 
peak near 0.3 \AA$^{-1}$.  
This feature vanishes in 
the nematic phase below $T_{NI}^{0}$.
Careful measurements also revealed evidence of this peak at high temperatures 
in 8CB confined in
aerosil gels.  
 We note that this feature
introduces a small inconsistency in our treatment of the background in 
Eq.~(2).
Specifically, $B(q)$ is obtained from the scattering intensity above 
$T_{NI}$, but any high-temperature liquid crystal contribution should not be included 
as part of the background below $T_{NI}$.  However, we have found 
that the liquid crystal contributions to $B(q)$ are sufficiently small 
that including or excluding them does not influence the characterization 
of the 
smectic peaks at $q \approx 0.2$ \AA$^{-1}$.  The relative strengths of the 
aerosil and liquid crystal scattering can, however, provide a method for
normalizing the scattering intensity and thus for comparing the intensities 
from different samples.  In particular, we decompose $B(q)$ into two separate contributions:

\begin{equation}
I(q, T>T_{NI}^{0})=B(q)=v_{S}B_{S}(q)+v_{LC}B_{LC}(q)
\end{equation}
where the shape of the liquid crystal contribution, $B_{LC}(q)$, is obtained 
from measurements on 
pure 8CB and the aerosil gel contribution, $B_{S}(q)$, is assumed to have a simple 
power-law form~\cite{germano}.  The ratio of the prefactors 
$v_{LC}/v_{S}$ scales with aerosil volume fraction as 
expected~\cite{sungilthesis}.  The prefactor to the liquid crystal contribution, 
$v_{LC}$, which is proportional to the quantity of 8CB in the beam,
provides a method for placing the scattering intensities on an 
absolute scale.  We define  normalized 
strengths for the thermal and static fluctuation terms by
\begin{eqnarray}
    \sigma_{1}^{N} & = & \sigma_{1}/v_{LC} \\
    a_{2}^{N} & = & a_{2}/v_{LC}
    \end{eqnarray}
which give the sizes of the two terms for each sample normalized by the scattering volume of 
8CB for that sample.

\begin{figure}
\centering\includegraphics[scale=0.8]{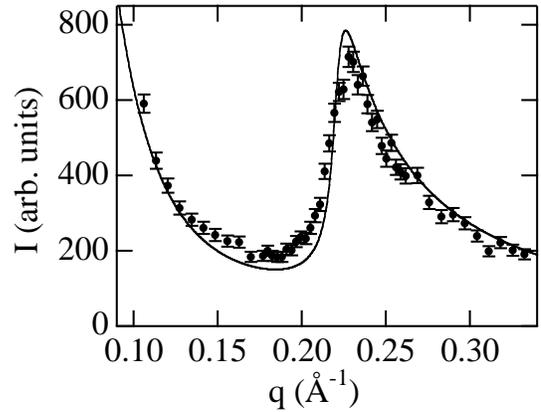}
\caption{Smectic peak measured for $\overline{8}$S5 confined in an 
aerosil gel with $\rho_{S}=0.030$ g/cm$^{3}$ at 339.04 K, 2.56 K {\it above} the N-SmA transition 
temperature of pure $\overline{8}$S5.  The solid line is a fit 
with Eqs.~(1) and (2), using 
values of $\xi_{\bot}(\xi_{\|})$ and $c(\xi_{\|})$ from 
pure $\overline{8}$S5 and with $a_{2}=0$, so that only the thermal 
fluctuations are 
included in describing the smectic correlations.  The strongly 
asymmetric lineshape is a consequence of the powder averaging.}
\label{fig4}
\end{figure}

\begin{figure}
\centering\includegraphics[scale=0.8]{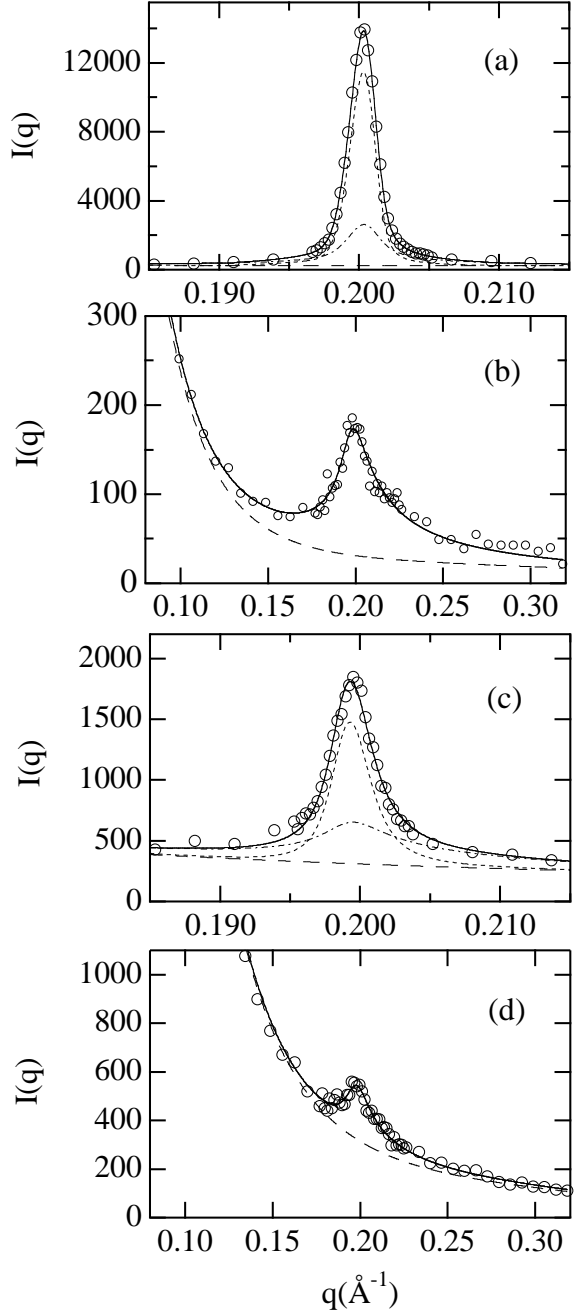}
\caption{Scattering intensity, $I(q)$, for 8CB in confined aerosil gels. 
(a) $\rho_{S}=0.025$ g/cm$^{3}$ at 288.5 K, 
(b) $\rho_{S}=0.025$ g/cm$^{3}$ at 307.6 K, (c) $\rho_{S}=0.282$ 
g/cm$^{3}$ at 300.6 K, and (d) $\rho_{S}=0.282$ g/cm$^{3}$ at 306.6 
K.  
The solid line in each panel is a fit with 
Eqs.~(1) and (2), and the dashed line is 
the background contribution from the aerosil gel structure.  In panels 
(a) and (c), the dash-dotted line is the contribution from the first term in 
Eq.~(1), representing thermal fluctuations, plus the background,
and the dotted line is the second term, 
representing static fluctuations, plus the background.  In panels (b) 
and (d), the temperatures are above a pseudotransition temperature $T^{*}$ 
defined in Sec. IIID, and the smectic correlations are described entirely by thermal fluctuations.}
\label{fig5}
\end{figure}

Figure 5, which shows representative fits to the 
lineshapes of two samples with Eqs.~(1) and (2), illustrates the good agreement we observe 
for this fitting form over the entire investigated range of densities and 
temperatures.  The solid lines in the figure correspond to the full fit 
while the dotted, dash-dotted and dashed lines correspond to contributions made by 
each of the three terms -- thermal fluctuations, static fluctuations, 
and background -- respectively.  As mentioned above, in the fits with Eq.~(1),  
$\xi_{\bot}$ and $c$ are treated as functions of 
$\xi_{\|}$, with $\xi_{\bot}(\xi_{\|})$ and $c(\xi_{\|})$ 
set by their relations in pure 8CB.  Thus, each fit has only five free parameters -- 
$q_{0}$, $A(T)$, $\xi_{\|}$, $\sigma_{1}$, and $a_{2}$ -- with two 
parameters, $\xi_{\|}$ and $a_{2}/\sigma_{1}$, controlling the 
{\it shape} of the peaks.

\subsection{Smectic layer spacing}

The wavevector $q_{0}$ in Eq.~(1) characterizes the spacing of the 
smectic layers, and Fig.~6 displays the values of $q_{0}$ extracted from fits
for three aerosil densities.  At high temperatures, where the smectic peak 
has a large width and small amplitude, the scatter in $q_{0}$ is too 
large to determine confidently any systematic dependence on 
temperature or $\rho_{S}$.  At low temperatures, $q_{0}$ appears to 
be virtually independent of $\rho_{S}$ and increases slightly with 
decreasing temperature.  This temperature dependence is shown in 
Fig.~6 
as a function of the difference in temperature from $T^{*}$, the 
temperature below which the magnitude of static smectic fluctuations
becomes non-zero.  The 
procedure for obtaining $T^{*}$ and its value as a function of 
$\rho_{S}$ are discussed below in Sec. IIID.

The temperature dependence for $q_{0}$ was also measured for pure 8CB below 
$T_{NA}^{0}$.  Due to small systematic uncertainties 
introduced by variations in the x-ray optics, precise comparisons on an absolute scale
between the values of $q_{0}$ 
for pure 8CB and those for the aerosil samples should not 
be made.  
However, $q_{0}$ for the aerosil samples at 
low temperatures and for pure 8CB in the smectic phase
showed very similar temperature 
dependencies, and we estimate that for a given effective temperature 
difference, $T-T^{*}$, the 
fractional difference in $q_{0}$ between pure 8CB and any aerosil sample is 
not more than 0.2\%.   Thus, the presence of the aerosil has little, 
if any, effect on the partial bilayer smectic layer spacing in 8CB.

\begin{figure}
\centering\includegraphics[scale=0.8]{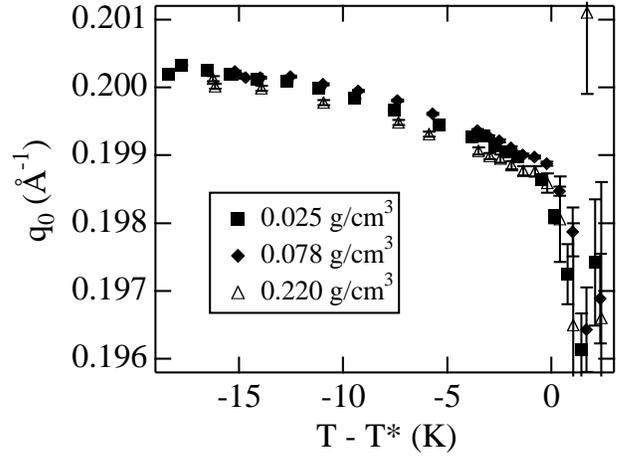}
\caption{The ordering wavevector, $q_{0}$, extracted from the fits with Eqs.~(1) and (2)
for smectic correlations in 8CB confined in 
aerosil gels with three $\rho_{S}$ values as specified.  
Temperature is plotted as the difference from $T^{*}$, the 
onset temperature for static fluctuations.  Values for $T^{*}$ are 
given in the inset on Fig.~8.}
\label{fig6}
\end{figure}

\subsection{Thermal fluctuations}

The amplitude of the thermal fluctuation term, 
$\sigma_{1}^{N}$, is displayed Fig.~7 as a function of temperature for 
several values of $\rho_{S}$.  
Consistent with Fig.~3, the 
temperature dependence of $\sigma_{1}^{N}$ at temperatures well above 
$T^{*}$ tracks that of the susceptibility for pure 8CB, which is shown 
by the solid line in the figure.  However, unlike the pure system, 
where the susceptibility diverges at $T_{NA}^{0}$, $\sigma_{1}^{N}$ 
for 8CB confined in aerosil gels remains finite through the ``transition'' region.  

At temperatures below $T^{*}$, $\sigma_{1}^{N}$ exhibits a roughly 
temperature independent value.  In Fig.~7, $\sigma_{1}^{N}$ for the 
$\rho_{S}=0.220$ and $\rho_{S}=0.282$ samples are shown 
as free parameters to low temperatures in order to 
illustrate the very weak temperature dependence 
at low $T$.  Below the pseudo-transition region near $T^{*}$, the 
thermal term should be dominated by long wavelength excitations, and 
the scattering intensity from these contributions should be dictated by the 
Bose occupation factor, which gives $\sigma_{1}^{N}(T) \propto T$.  
Over the narrow absolute temperature range below $T^{*}$ (approximately 290 K to 
307 K), 
the roughly constant behavior for $\sigma_{1}^{N}$ is 
consistent with this expectation.  On the basis of this observed low-temperature 
$\sigma_{1}^{N}$ 
behavior in unconstrained fits, we have repeated the fitting with Eqs.~(1) 
and (2) for each $\rho_{S}$ for $T<T^{*}$ with $\sigma_{1}^{N}$ fixed at 
the average low temperature value for that density.  
This second iteration of fitting was designed to obtain better stability in 
the results for the other fit parameters (namely $a_{2}^{N}$ and 
$\xi_{\|}$) that are coupled to $\sigma_{1}^{N}$~\cite{paramcoupling}.  
The results for $\rho_{S}=0.025, 0.041, 0.051,$ and $0.105$ 
in Fig.~7 show $\sigma_{1}^{N}$ held constant at low temperature 
under this procedure.
Near and above $T^{*}$, we treat $\sigma_{1}^{N}$ as 
a free, temperature-dependent parameter, leading to the behavior shown in Fig.~7.

\begin{figure}
\centering\includegraphics[scale=0.83]{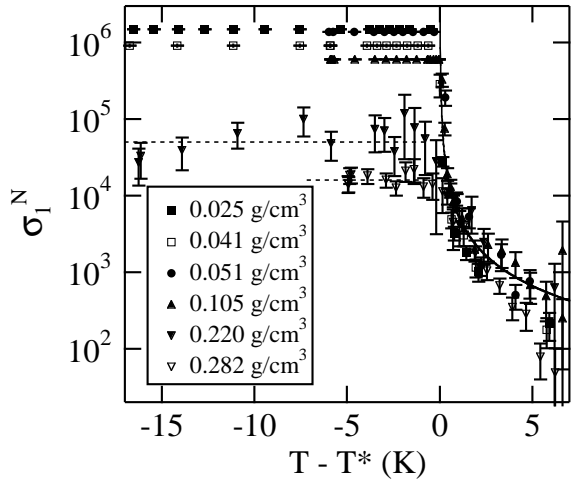}
\caption{ Amplitude of the thermal fluctuations, $\sigma_{1}^{N}$, for 8CB confined in 
aerosil gels of several densities $\rho_{S}$ as specified.
Temperature is plotted as the difference from $T^{*}$, the 
onset temperature for static fluctuations.  Values for $T^{*}$ are 
given in the inset on Fig.~8.  The solid line is the temperature 
dependence of the susceptibility
for pure 8CB, with $T^{*}($pure$)=T_{NA}^{0}$.  }
\label{fig7}
\end{figure}

\subsection{Static fluctuations}

As mentioned above, the scattering intensity from 8CB in aerosil gels above $T_{NA}^{0}$ 
is well
described by thermal fluctuations alone, and the static term in 
Eq.~(1) is effectively zero.  With decreasing temperature, the 
contributions from static fluctuations rise sharply from zero.
Figure 8 displays the 
temperature dependence of the static fluctuation term, 
$a_{2}^{N}$, for three values of $\rho_{S}$.  The solid lines 
in the figure are the results of fits with the form 

\begin{equation}
    a_{2}^{N} = B(T^{*}-T)^{x}, 
\end{equation}
where $B$, $x$, and $T^{*}$ depend on $\rho_{S}$.    Values 
for $B$, $x$, and $T^{*}$ are tabulated in Paper II.  As the figure 
illustrates, this expression describes the temperature dependence 
of $a_{2}^{N}$ quite well.  The significance of this functional form is discussed
in Section IV.  

Fits with Eq.~(6) provide unambiguously the onset temperature $T^{*}$
of static fluctuations induced by the quenched disorder.  The 
open symbols 
in the inset of 
Fig.~8 are the resulting $T^{*}$ values as a function of $\rho_{S}$.  
As the inset illustrates, the onset of static smectic order becomes 
increasingly suppressed in temperature as $\rho_{S}$ increases.  This 
suppression in $T^{*}$ 
quantitatively tracks the decreasing 
position of the peak in the heat capacity with $\rho_{S}$
observed in calorimetry studies~\cite{germano} and shown in the inset with 
solid 
symbols.  Due to the considerable uncertainty in the values of $T^{*}$ 
extracted from the fits with Eq.~(6), the nonmonotonic variation in 
``transition'' temperature with $\rho_{S}$ that was observed 
calorimetrically for 
$\rho_{S}<0.1$~\cite{germano} cannot be resolved.

\begin{figure}
\centering\includegraphics[scale=0.8]{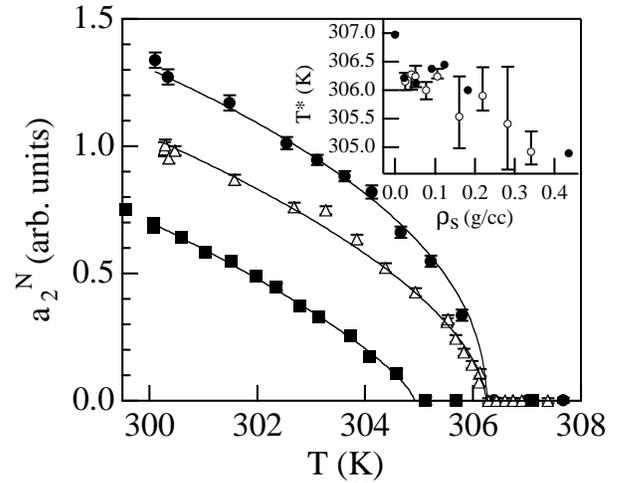}
\caption{The integrated area, $a_{2}^{N}$, of the static fluctuation term in $S(q)$ for
three values of $\rho_{S}$: 0.051 (solid circles), 0.105 (open 
triangles), and 0.341 (solid squares).  The values for 
$\rho_{S}=0.051$ g/cm$^{3}$ and $\rho_{S}=0.105$ g/cm$^{3}$ have been multiplied by 
7 and 5, respectively, for clarity.  The solid lines 
are fits with Eq.~(6).  The open circles in the inset are the values of $T^{*}$, the onset 
temperature for static fluctuations extracted 
from these fits, as a function of $\rho_{S}$.  The solid circles 
in the inset are the peak positions in the specific heat as obtained 
from calorimetry studies [7].}
\label{fig8}
\end{figure}

\subsection{Smectic correlation length}

Consistent with Fig.~3, the values of
the smectic correlation lengths, $\xi_{\|}$, 
extracted from the fits with Eqs.~(1) and (2) for 
$T>T^{*}$ are like 
those of pure 8CB and track its temperature dependence.  
Figure 9(a) displays  $\xi_{\|}$ at temperatures above $T^{*}$.  Any 
possible systematic variation in $\xi_{\|}$ with $\rho_{S}$ 
is overwhelmed by the scatter in the data.  The solid line in the figure is 
$\xi_{\|}$ for pure 8CB~\cite{ocko}, where we use $T_{NA}^{0}$ for $T^{*}$.  
A detailed scaling analysis of the temperature dependence expected for 
$\xi_{\|}$ of 8CB + aerosils above 
$T^{*}$ is provided in Paper II. 

As with $\sigma_{1}^{N}$, $\xi_{\|}$ for aerosil samples fails to track 
the diverging behavior of the pure 
system through $T_{NA}$ and remains finite to low temperature.  
At temperatures below $T^{*}$, $\xi_{\|}$ approaches an essentially 
temperature independent value.  Figure 9(b) shows the correlation 
lengths at low temperature for four densities of aerosil.  The 
variation of the low temperature correlation length, $\xi_{\|}^{LT} \equiv 
\xi_{\|}(T<<T^{*})$, with $\rho_{S}$ is described in Section IV.

\section{Discussion}

\begin{figure}
\centering\includegraphics[scale=0.8]{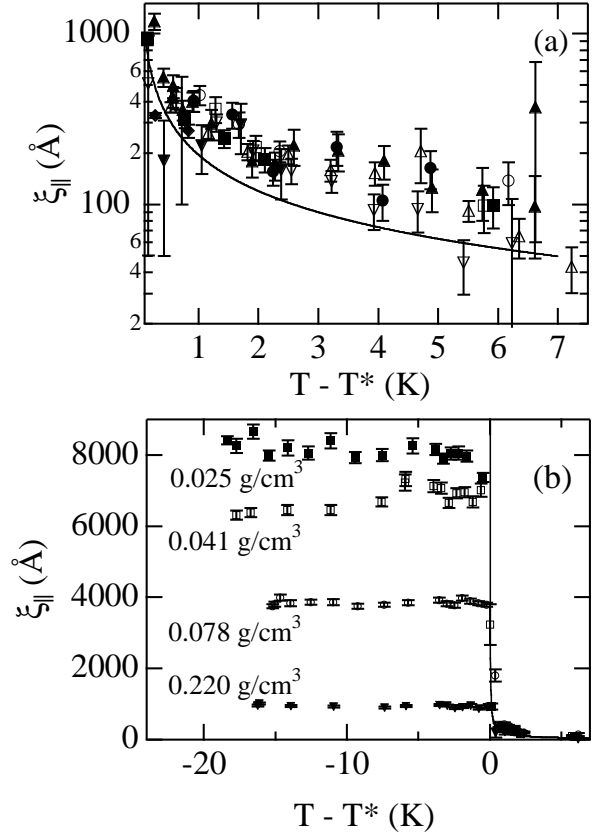}
\caption{The smectic correlation length, $\xi_{\|}$, for 8CB confined in 
aerosil gels.  (a)  $\xi_{\|}$ for $T>T^{*}$ compared with 
that of pure 8CB [solid line with $T^{*}($pure$)=T_{NA}^{0}$] for 
values of $\rho_{S}$ as specified in Fig.~10. (b) 
$\xi_{\|}$ values given to low temperatures for four values of $\rho_{S}$.}
\label{fig9}
\end{figure}

As the results outlined in Sec. III demonstrate, confinement by aerosil gels
strongly affects the smectic transition in 8CB.  
The success of Eq.~(1) indicates that these effects can be modeled as
random fields.  Consistent with 
theoretical predictions the random fields destroy the transition to a 
quasi-long-range-ordered smectic phase and replace it with the formation 
of short-ranged correlations.  In this Section, 
we investigate the relationship between the strength of this disorder
and the corresponding short-ranged smectic order, including possible 
connections with theoretical predictions, as well as aspects of 
the pseudocritical behavior exhibited by 8CB confined in aerosil gels.

\subsection{Pseudocritical behavior}

\begin{figure}
\centering\includegraphics[scale=0.8]{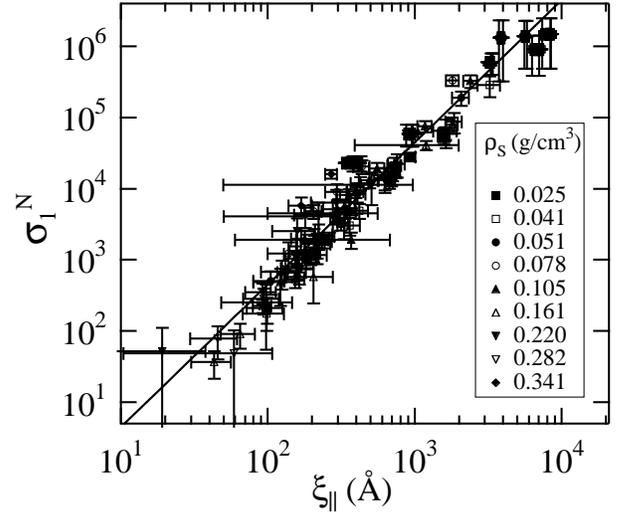}
\caption{ Amplitude of the thermal fluctuations, $\sigma_{1}^{N}$, 
versus correlation length $\xi_{\|}$ for 8CB confined in 
aerosil gels.  The plotted values span the full experimental range of 
$\rho_{S}$ and temperature.  The solid line is the relation 
$\sigma_{1}^{N} \propto \xi_{\|}^{2}$.}
\label{fig10}
\end{figure}

As Figs.~7 and 9(a) illustrate, the smectic scattering in 8CB confined in 
aerosil gels above $T^{*}$ resembles, to a first approximation, 
thermal critical fluctuations unaffected by quenched disorder.  
This close connection between the thermal term in Eq.~(1) for 8CB + 
aerosils and the corresponding term for pure liquid crystals in the 
nematic phase is strengthened by considering the scaling relation 
between $\sigma_{1}^{N}$ and $\xi_{\|}$.  In a pure isotropic critical 
system for $T>T_{c}$, one has $\sigma_{1} \propto t^{-\gamma}$ and 
$\xi \propto t^{-\nu}$ which leads to $\sigma_{1} \propto 
\xi^{\gamma/\nu} = \xi^{2-\eta}$.  For liquid crystals, which exhibit 
critical anisotropy in the behavior of the correlation lengths, one 
obtains $\sigma_{1} \propto 
\xi_{\|}^{\gamma/\nu_{\|}}$ where $\gamma/\nu_{\|} = 2 - \eta_{\|}$.  
This scaling relation between the thermal susceptibility and the 
parallel correlation length for 8CB + aerosils can be written as 
\begin{equation}
\sigma_{1}^{N} = A\xi_{\|}^{2-\eta}
\end{equation}
where the subscript on $\eta$ has been dropped for convenience.  
Figure 10 displays a log-log plot of $\sigma_{1}^{N}$ versus $\xi_{\|}$ for a series of samples 
of varying $\rho_{S}$ and a line representing $\sigma_{1} \propto 
\xi_{\|}^{2}$.
As the figure demonstrates, the scaling relation holds for 8CB confined in aerosil 
gels over the entire experimental temperature 
range, both in the pseudo-critical regime 
above $T^{*}$ and at low temperature where both $\sigma_{1}^{N}$ and 
$\xi_{\|}$ have only weak 
temperature dependence.  Further, the scaling amplitude, $A$, shows 
no systematic variation with $\rho_{S}$.  Indeed, the successful 
collapse of data from different samples in Fig.~10 suggests the validity of
the normalization procedure for $\sigma_{1}^{N}$ using $v_{LC}$.  Also 
clear from the figure is the fact that $2-\eta$ is close to 2 for 8CB + aerosils.  The mean field 
and the Gaussian tricritical values of $2-\eta$ are 2.00, and the 3D-XY 
value is 1.962.  Furthermore, the value of $\gamma/\nu_{\|}$ is 1.88 
for pure 8CB~\cite{ocko}.  The correspondence between the scaling 
given in Eq.~(7) applied to 8CB + aerosil gels of varying $\rho_{S}$ 
and that for pure liquid crystals is discussed below.

This pseudocritical behavior of 8CB confined in aerosil gels also extends below $T^{*}$, 
where $c$ in Eq.~(1) approaches a small constant value, 
and $a_{2}$ 
becomes proportional to the integrated 
intensity of the static fluctuation term in $I(q)$.  The growth of 
this intensity, shown in Fig.~8, strongly resembles 
that of an order parameter squared, as the success of the fits of 
$a_{2}$ with 
Eq.~(6) illustrates.  Figure 11(a) shows 
the values of the effective exponent $x$ extracted 
from these fits as a function of $\rho_{S}$.  For pure liquid crystals,
the critical behavior observed at the N-SmA transition is
affected by the proximity of the higher 
temperature nematic to isotropic transition through a coupling between 
the nematic and smectic order parameters
({\it e.g.} de Gennes coupling; see Paper II)~\cite{nounesis}.  The strength of this coupling 
depends on the magnitude of the nematic susceptibility at $T_{NA}$, 
which for pure liquid crystals depends on the width of the nematic 
range and can be roughly parameterized by the MacMillan ratio, 
$R_{M} \equiv T_{NA}/T_{NI}$.  When $R_{M}$ becomes small ($R_{M} 
\approx 0.7$), the N-SmA transition is observed to approach 3D-XY 
behavior but still exhibits a small anisotropy in the critical 
correlations~\cite{young}, while it approaches tricritical behavior and 
then becomes first order as 
$R_{M}\rightarrow 1$~\cite{nounesis}.  Values of $2\beta$, 
determined through the Rushbrooke scaling equality $2\beta = 2-\alpha-\gamma$~\cite{beta}, for a 
series of pure liquid crystals~\cite{nounesis} with varying $R_{M}$ are included 
in Fig.~11(a).   The effective 8CB + aerosil exponents, $x$, span the range from 
near the tricritical value $2\beta = 0.5$ (like 
pure 8CB) at small $\rho_{S}$ to 3D XY-like, $2\beta = 0.691$, at large 
$\rho_{S}$.  With
increasing $\rho_{S}$, the quenched disorder has the apparent effect 
of suppressing the nematic susceptibility and thus the 
coupling that drives 
the critical behavior of the N-SmA transition
in pure liquid crystals away from 3D-XY universality.

\begin{figure}[top]
\centering\includegraphics[scale=0.8]{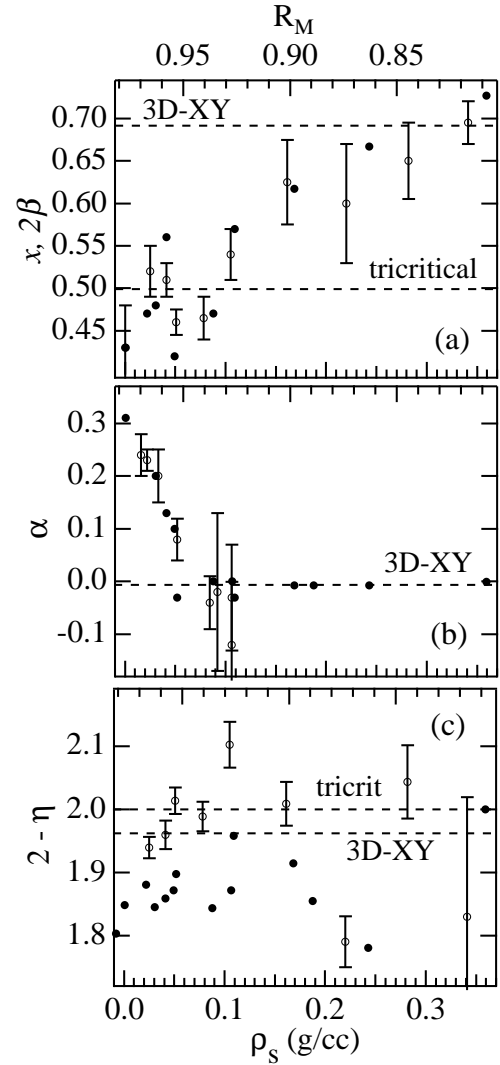}
\caption{(a) The effective exponent $x$ from fits to $a_{2}(T)$ with 
Eq.~(6) for 8CB+aerosil samples (open circles)
versus $\rho_{S}$.  Also given are values of the squared order parameter 
exponent, $2\beta$ (filled circles), for a series of pure 
liquid crystals with varying MacMillan ratios, $R_{M} \equiv 
T_{NA}/T_{NI}$, as determined from data given in Ref.~[34].  The
effective exponent $x$ spans the 
range from near a tricritical value, $2\beta=0.5$, at small $\rho_{S}$ to 
a 3D XY-like value, $2\beta=0.691$, at large $\rho_{S}$.   (b) the 
effective exponent, $\alpha$, from calorimetric 
studies on 8CB confined in aerosil gels [7] (open circles) 
along with values of $\alpha$ (filled circles) for a series of pure 
liquid crystals with varying $R_{M}$, as given in Ref.~[34].  
(c) the effective exponent $2-\eta$  obtained from 
fits with the form $\sigma_{1}^{N} = A\xi_{\|}^{2-\eta}$ versus 
$\rho_{S}$ (open circles) and the critical exponent
$\gamma/\nu_{\|}$ ($=2-\eta_{\|}$)  for a series of pure 
liquid crystals as a function of $R_{M}$.  Note that the linear
scaling between $\rho_{S}$ and $R_{M}$, given by Eq.~(8),
is the same for all three plots.}
\label{fig11}
\end{figure}

This trend agrees very closely with the corresponding behavior 
observed in specific heat studies of 8CB with dispersed 
aerosils~\cite{germano}.  Figure 11(b) shows the values of the effective heat capacity 
exponent $\alpha$ 
from the calorimetric study versus $\rho_{S}$.  As the figure illustrates, $\alpha$ 
decreases toward the 3D-XY value with increasing 
$\rho_{S}$.  The solid circles in Fig.~11(b) are 
the critical exponents measured for pure liquid crystals as a 
function of $R_{M}$.  
The same linear scaling between $\rho_{S}$ and $R_{M}$ 
successfully collapses the aerosil data onto the pure liquid crystals in 
Figs.~11(a) to 11(b) and suggests the notion of an effective 
MacMillan ratio
for 8CB in aerosil gels,

\begin{equation}
    R_{M}^{eff}(sil)=0.977-0.47\rho_{S}
\end{equation}
This general relationship between $R_{M}^{eff}(sil)$ and $\rho_{S}$ 
for two distinct pseudo-critical behaviors strengthens the idea that 
the presence of quenched disorder suppresses the nematic susceptibility 
in 8CB near $T^{*}$.  This suppression likely results from the role of 
surface anchoring on the gel strands in changing the properties of the 
nematic since the observed ratio $T_{NA}/T_{NI}$ for 8CB + aerosils 
varies little with $\rho_{S}$~\cite{germano}.   Figure 11(c) displays values of $2-\eta_{\|}$ 
extracted from fits with Eq.~(7) to $\sigma_{1}^{N}$ and 
$\xi_{\|}$ values obtained for different $\rho_{S}$ samples.  Also 
shown in Fig.~11(c) are 
values of $\gamma/\nu_{\|}$  ($=2-\eta_{\|}$)
for pure liquid crystals as a function of $R_{M}$.  Due to the small 
difference in $2-\eta_{\|}$ between the tricritical and 3D-XY values and 
the large scatter in the results both for pure liquid crystals and 
for 8CB + aerosil samples, systematic trends like those seen in Fig. 11(a) 
and 11(b) are less clear for $2-\eta$.  Tabulated values for 
$2-\eta$ are given in Paper II.

\subsection{Low temperature correlations}

\begin{figure}[top]
\centering\includegraphics[scale=0.8]{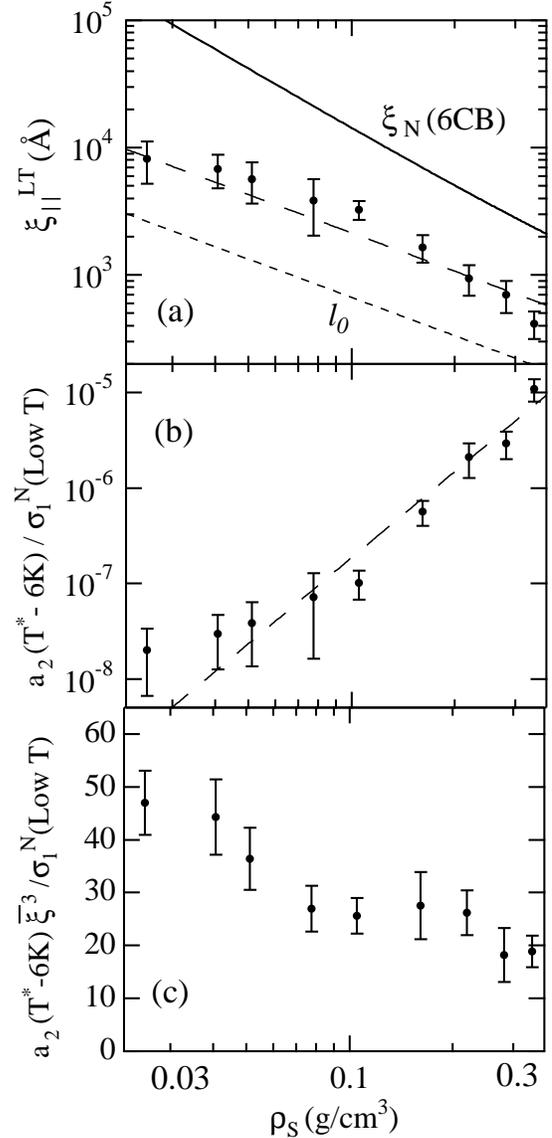}
\caption{(a) Low temperature correlation length, $\xi_{\|}^{LT}$, as a 
function of $\rho_{S}$.  The solid line is the nematic 
correlation length (for 6CB) measured by static light scattering [10], and the 
dashed line is the mean aerosil void size, $l_{0}$.  The long-dashed line 
represents the form $\xi_{\|}^{LT} \propto \rho_{S}^{-1}$.
(b)  The ratio $a_{2}^{N}/\sigma_{1}^{N}($low T$)$ at the fixed 
temperature difference $T^{*}-T = 6$ K as a function of $\rho_{S}$.  
The long-dashed line is the relation $a_{2}^{N}/\sigma_{1}^{N}$(low T)$ \propto 
\rho_{S}^{3}$.  (c)  The ratio 
$a_{2}^{N}\overline{\xi}^{3}/\sigma_{1}^{N}($low T$)$ at the fixed 
temperature difference $T^{*}-T = 6$ K as a function of $\rho_{S}$.}
\label{fig12}
\end{figure}

As Fig.~9(b) demonstrates, the correlation length for smectic order in 
8CB confined in aerosil gels saturates at low-temperature.  
Figure 12(a) shows these low temperature 
parallel correlation lengths,
$\xi_{\|}^{LT}$,
as a function of $\rho_{S}$.  The solid 
line in Fig.~12(a) shows the trend in the {\it nematic} correlation lengths of the 
homolog 6CB 
with dispersed aerosils, as measured by Bellini 
\et ~\cite{bellini-sil-PRE} with static light scattering.  These light 
scattering studies demonstrate that the 
nematic order in a liquid crystal with dispersed aerosil breaks up into very 
large, but finite, domains.  Since smectic layering forms within 
regions with nematic order, the nematic domain sizes set upper limits 
for the range of the smectic correlations.  
However, $\xi_{\|}^{LT}$ remains 
well below these limits for all $\rho_{S}$, demonstrating the  
sensitivity of the smectic phase to quenched disorder.  The dashed line 
in Fig.~12(a) is the mean aerosil 
void size, $l_{0}$, as a function of $\rho_{S}$~\cite{germano,paperII}.  
Over the range of densities studied, $\xi_{\|}^{LT}$ exceeds
$l_{0}$ by a factor of three or more, 
consistent with an intermediate strength disordering field.  Paper II 
provides a detailed discussion of the role of $l_{0}$ in setting the 
range of the smectic correlations.

Figure 12(b) displays $a_{2}^{N}/\sigma_{1}^{N}$ as a function of $\rho_{S}$
for $T = T^{*}-6$ K.  
We plot this ratio rather than $\sigma_{1}^{N}$ or 
$a_{2}^{N}$ alone to characterize the magnitude of the 
fluctuations in order to remove any errors that the uncertainties in 
$v_{LC}$ might introduce through the normalization.  Because 
the exponent, $x$, characterizing the growth of $a_{2}^{N}$ below 
$T^{*}$ is only weakly dependent on $\rho_{S}$, the trends observed 
for this ratio are insensitive (on a logarithmic scale) 
to the temperature difference $T^{*}-T$ used to evaluate $a_{2}^{N}$.  Figure 12(c) 
shows the quantity $a_{2}^{N}\overline{\xi}^{3}/\sigma_{1}^{N}$, where 
$\overline{\xi} \equiv (\xi_{\|}\xi_{\bot}^{2})^{1/3}$ is the mean 
correlation length at low temperature.  This 
quantity, which is approximately the ratio of the integrated intensities of the 
thermal and disorder terms at low temperature, shows little variation with 
$\rho_{S}$.  

\subsection{Comparisons with theory}

As mentioned in the introduction, random field models have served as 
important examples for understanding possible consequences of quenched 
disorder, and extensive theoretical work has been devoted to random 
field systems.  For the case of transitions that break a continuous 
symmetry in three dimensions, like the nematic to smectic-A transition, the domain 
wall energy arguments of Imry and Ma conclude that the
transition is unstable to arbitrarily weak 
random fields~\cite{imry-ma}.  Subsequent theoretical work has 
rigorously established this conclusion~\cite{aizenman}, consistent with the 
short-ranged smectic order measured for all 8CB + aerosil samples well below $T_{NA}^{0}$.
For quantitative comparisons with these short-range smectic 
correlations two theoretical efforts provide guidance.  In 
the first, Aharony and Pytte\cite{aharony&pytte} apply scaling arguments to predict the low temperature form of 
the structure factor for systems with transitions breaking continuous 
symmetries in the presence of random fields.  In the second, Radzihovsky and 
Toner\cite{leo} develop a detailed theory for smectics with random quenched disorder.  

The strength of the disorder imposed on smectic 8CB by confinement in 
aerosil gels clearly correlates with the density of the aerosil; 
however, the precise relation between density and disorder is 
unknown.  Within the simplest assumption, confinement in the gels
induces random fields whose disorder strength is proportional to 
the gel density.  With this assumption, the scaling arguments of Aharony and 
Pytte  predict $\xi \propto 
\Delta^{-1} \propto \rho_{S}^{-1}$, where $\Delta$ is the strength of the 
random field disorder.  The exponent $-1$ is obtained from $-1/(d_{c}-d)$ 
where $d=3$ is the spatial dimension and $d_{c}=4$ is the lower critical 
dimension for the random field XY model~\cite{lcd}.  The 
long-dashed line in Fig.~12(a) is the power-law form $\xi_{\|}^{LT} \propto 
\rho_{S}^{-1}$.  Over the range of $\rho_{S}$ covered 
experimentally, measured values of $\xi_{\|}^{LT}$ appear roughly to 
follow this behavior.  A best fit to a 
power-law form over the full range of $\rho_{S}$ gives $\xi_{\|}^{LT} \propto 
\rho_{S}^{-1.1 \pm 0.2}$.
The scaling arguments of Aharony and Pytte further lead to 
$a_{2}^{N}/\sigma_{1}^{N} \propto 
\Delta^{3}$.  The long-dashed line in Fig.~12(b) is this 
relation, again assuming $\Delta \propto \rho_{S}$.  As
with $\xi_{\|}$, $a_{2}^{N}/\sigma_{1}^{N}$ follows a trend crudely 
consistent with the predicted power-law behavior, but clear deviations 
from the predicted form are also apparent.  One possible source for 
these discrepancies could be a more subtle relation between 
$\rho_{S}$ and disorder strength.  Paper II elaborates on this issue.  

In order to make comparisons that are free of the dependence of 
disorder on $\rho_{S}$, we plot 
in Fig.~13 $a_{2}^{N}/\sigma_{1}^{N}$ against $\overline{\xi}$.  
In this plot we use $\overline{\xi}$ rather than $\xi_{\|}$
since the scaling theory is for an isotropic system, and 
$\overline{\xi}^{3}$ represents the correlation volume.  
According to the scaling theory, $a_{2}^{N}/\sigma_{1}^{N}$
should vary with correlation length like $\overline{\xi}^{-3}$, while a 
best fit shown by the solid line gives $\overline{\xi}^{-2.68 \pm 
0.26}$.  This relatively good agreement with the theory is also 
reflected in the weak variation with disorder strength of 
$a_{2}^{N}\overline{\xi}^{3}/\sigma_{1}^{N}$ shown in Fig.~12(c).

A second possible source of discrepancy between the smectic correlations 
in 8CB + aerosil and the scaling arguments of Aharony and 
Pytte is that additional disordering effects enter the confined smectic 
beyond those of the random field 3D XY system.  In their discussion of smectic 
liquid crystals with quenched disorder, Radzihovsky and Toner identify 
two sources of disorder, tilt disorder that couples to
orientation (ie, the nematic director) and a layer displacement 
disorder that couples directly to the smectic order parameter.  While 
the latter corresponds to the disorder of the XY model, 
the former is predicted to be the dominant disorder for
the smectic.  An important prediction of the work 
of Radzihovsky and Toner is that the disordered smectic should possess 
anomalous 
elasticity, and one consequence of this prediction is that the 
correlation length ratio, $\xi_{\|}/\xi_{\bot}$, is reduced from that 
of the pure system.  In our modeling of the smectic correlations with 
Eq.~(1), we 
find that maintaining $\xi_{\|}/\xi_{\bot}$ equal to its values in pure 
8CB produces good fits to the data.  However, the 
powder nature of the measured lineshapes creates considerable 
uncertainties in $\xi_{\bot}$, so that we can not exclude some reduction 
in the correlation length ratio.  Despite this inconsistency, the 
prediction from Radzihovsky and Toner that $\xi_{\|}^{LT} \propto 
\Delta^{-1}$ is roughly consistent with our results in 
Fig.~12(a), assuming again $\Delta \propto \rho_{S}$ holds 
approximately.

\begin{figure}
\centering\includegraphics[scale=0.8]{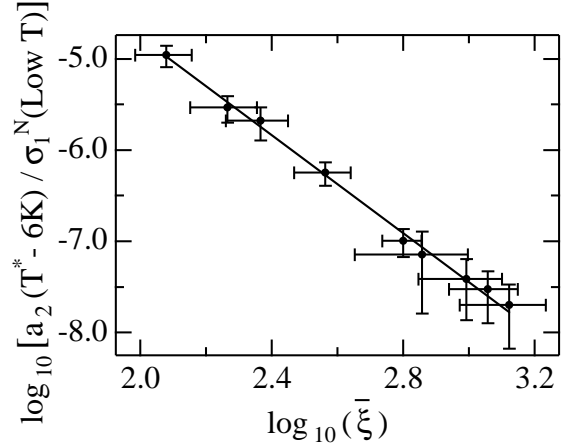}
\caption{The ratio $a_{2}^{N}/\sigma_{1}^{N}($low T$)$ at the fixed 
temperature difference $T^{*}-T = 6$ K as a function of $\overline{\xi}$.
The solid line is a power law fit 
giving exponent value of -2.68.}
\label{fig13}
\end{figure}

Radzihovsky and Toner further 
predict that $\xi(T) \propto B(T)^{1/\Gamma}$ at low temperature, where 
$B(T)$ is the layer compression modulus of the pure liquid crystal.  
Within this theory, 
$1/\Gamma > 1/2$ is required for a smectic Bragg glass phase to be 
stable.  Published data for $B(T)$ of 8CB extend only 3 K below 
$T_{NA}^{0}$~\cite{BofT}, so a direct test of this scaling prediction is difficult.  
However, the essentially temperature independent behavior of 
$\xi_{\|}$ at low temperature shown in Fig.~9(b) indicates that such 
scaling does not hold for 8CB in aerosils.  
We note that this conclusion differs from that reported for 8CB confined in 
aerogels, in which the analysis of those x-ray scattering results yielded 
$\xi(T)$ consistent with $\xi(T) \propto B(T)^{1/\Gamma}$ and with 
marginal Bragg glass stability~\cite{clark-science}.  This 
contrast between the aerosil and aerogel results is somewhat 
surprising since the validity of the prediction of a Bragg glass is within 
a regime of weak disorder, and 
the aerosil gels impose a more gentle perturbation on the smectic than aerogels.  
One source for the different $\xi_{\|}(T)$ behaviors could be the 
partial
compliance of the aerosil gel to the elasticity of the liquid 
crystal, a feature which is not 
incorporated into the theory and which may complicate comparisons
with it.  However, a number of differences also exist between the 
approach used for analyzing the smectic lineshapes in the aerogel 
study\cite{rappaport,clark-science} 
and the analysis presented here for aerosil samples.  For 8CB + 
aerosil data, analysis like that used in
the aerogel study has strong consequences for $\xi(T)$.
Further study of other liquid crystals would clearly help in 
determining whether the contrast between the aerogel and aerosil 
systems in their agreement with Bragg glass stability is a consequence 
of elastic coupling that is not considered in the theory 
or, perhaps, an artifact of the differing approaches to the lineshape analysis.  
Such studies would also be invaluable in testing the generality of 
trends observed for 8CB, such as the crossover to 3D XY behavior shown 
in Fig.~11, with the introduction of quenched disorder through 
confinement in aerosil.\\
\\

\begin{centering}
    {\small \bf ACKNOWLEDGEMENTS\\}
\end{centering}

\vspace{0.25 in}

We thank S. Lamarra for technical support and P. Clegg and 
L. Radzihovsky for helpful discussions.  
This work was supported by the Natural Science and 
Engineering Research Council of Canada (Toronto) and by the NSF 
under CAREER Award No.~DMR-0134377 
(JHU), CAREER Award No.~DMR-0092786 (WPI), and Contract No.~DMR-0071256 
(MIT).\\
\\

\begin{centering}
    {\small \bf APPENDIX: ANALYTIC POWDER AVERAGING OF THE SMECTIC STRUCTURE 
FACTOR\\}
\end{centering}

\vspace{0.25 in}

The x-ray structure factor for the smectic fluctuations in 8CB 
confined by aerosil gels, given in Eq.~(1), includes two terms.  The 
first term (LFC) is an anisotropic Lorentzian with a 4th order
correction in the transverse direction that accounts for critical 
thermal fluctuations.  The second term (LFC$^{2}$) is proportional to 
the thermal term 
squared and accounts for random-field induced static 
fluctuations.
Although an anisotropic Lorentzian raised to a power along the transverse direction, 
$S({\bf q}) = {\sigma}/\left\{{\xi_{\parallel}^2
(q_{\parallel} - q_0)^2 + (1 + \xi_{\perp}^2 q_{\perp}^2)^{1 -
\eta_{\perp} / 2}}\right\}$, is equally successful in describing the
thermal fluctuations\cite{Ocko_Thesis}, we use the LFC because
its powder average can be calculated analytically.  The structure 
factor for a powdered sample
is equivalent to that for a single-domain sample
averaged with equal weight over all orientations of the smectic wave vector, ${\bf q}_0$, 
relative to the scattering wave vector, ${\bf q}$,
\begin{equation}
 S^{powder}(q) = \frac{1}{4\pi} \int d \Omega_{{\bf q}_0} \ S({\bf
 q}).
\end{equation}

For the LFC term, the integral has the form

 \begin{eqnarray}
\scriptstyle S_{LFC}^{powder}(q)  & = & \scriptstyle \frac{1}{4\pi} \int d \Omega \
    \frac{\sigma}{\left\{ 1 + \xi_{\parallel}^2 (q_{\parallel} - q_0)^2 + \xi_{\perp}^2 q_{\perp}^2
      + c \xi_\perp^4 q_\perp^4 \right\} }
    \nonumber \\
  & = & \scriptstyle \frac{1}{2} \int_{-1}^{1} d \cos \theta \
      \frac{\sigma}{ \left\{ 1 + \xi_{\parallel}^2 ( q \cos \theta - q_0 )^2 + \xi_{\perp}^2 q^2 \sin^2 \theta
      + c \xi_{\perp}^4 q^4 \sin^4 \theta
      \right\} } \nonumber \\
  & = & \scriptstyle \frac{\sigma}{2} \int_{-1}^{1} d \mu \
      \frac{1}{(A + 2 B \mu + C \mu^2 + D \mu^4)},
\end{eqnarray}

 where 
 \begin{eqnarray}
  A & = & 
  1 + \xi_{\parallel}^2 q_0^2 + \xi_{\perp}^2 q^2 + c \xi_\perp^4 q^4, \nonumber \\
  B & = & -\xi_{\|}^{2}q_{0}q,  \nonumber \\
  C & = &  (\xi_{\parallel}^2 - \xi_{\perp}^2) q^2 - 2 c \xi_\perp^4
  q^4,
  \nonumber \\
  D & = & c \xi_\perp^4 q^4 .
\end{eqnarray}
  This integral in turn can be written as 
  \begin{equation}
  S_{LFC}^{powder}(q) = \frac{1}{D} \int_{-1}^{1} d \mu \
       \prod_{i} \frac{1}{(\mu - \alpha_i)^m_i},
   \end{equation}
where $\alpha_i$ are the roots with multiplicity $m$ of the quartic equation:
\begin{equation}
    A + 2B\mu + C\mu^2 + D\mu^4 = 0.
\end{equation}
With the integrand in this form, it can be expanded into a sum of 
terms,
 \be
    \prod_{i} \frac{1}{(\mu - \alpha_i)^m_i} = \sum_{i, 1 \leq j \leq m }  \frac{ C_{ij} }{ (\mu -
      \alpha_i)^j},
 \ee
 where the coefficients $C_{ij}$ are obtained through the expansion
 process and are listed in Table I.  At this stage, each term can be 
 evaluated analytically using elementary definite
 integrals, such as
\begin{eqnarray}
\int_{-1}^{1} d\mu \frac{1}{\mu - \alpha} = \frac{1}{2}
\ln \left| \frac{ \left( 1 - \alpha_R \right)^2 + \alpha_I^2}{
\left( 1 + \alpha_R \right)^2 + \alpha_I^2 } \right| \nonumber\\
+ i \left[
 \tan^{-1} \left( \frac{1-\alpha_R}{\alpha_I} \right)
 + \tan^{-1} \left( \frac{1+\alpha_R}{\alpha_I} \right) \right], 
 \end{eqnarray}
 where $\alpha = \alpha_R + i \alpha_I $.

The analytic powder averaging of the LFC$^{2}$ term follows the same 
approach: 



 \ber
 \scriptstyle S_{LFC^2}^{powder}(q)  & = & \scriptstyle  \frac{1}{4\pi} \int d \Omega \
    \frac{\sigma}{\left\{ 1 + \xi_{\parallel}^2 (q_{\parallel} - q_0)^2 + \xi_{\perp}^2 q_{\perp}^2
      + c \xi_\perp^4 q_\perp^4 \right\}^2 }
    \nonumber \\
  & = & \scriptstyle \frac{\sigma}{2} \int_{-1}^{1} d \mu \
      \frac{1}{(A + 2 B \mu + C \mu^2 + D \mu^4)^2} \nonumber
      \\
  & = & \scriptstyle \frac{\sigma}{2 D^2} \int_{-1}^{1} d \mu \
      \prod_{i} \frac{1}{(\mu - \alpha_i)^{2m_i}}.
 \eer
 As with the LFC term, this integral can be expanded into a series of 
 elementary definite integrals.
 The coefficients for the expansion of the LFC$^2$ integral
 are listed in Table II.
 


\newpage

\onecolumn
\begin{table}[b]
\begin{tabular}{|c|c|}
 \hline
 Form of Expansion Term & Expansion Coefficients
 \\ \hline \hline



 $\frac{1}{(\mu - \alpha_1)(\mu - \alpha_2)(\mu - \alpha_3)(\mu -
 \alpha_4)}$
 & $C_{11} = \frac{1}{ (\alpha_1 - \alpha_2)(\alpha_1 - \alpha_3)(\alpha_1 -
 \alpha_4)}$ \\
 & $C_{21} = \frac{1}{(\alpha_2 - \alpha_1)(\alpha_2 - \alpha_3)(\alpha_2 -
 \alpha_4)}$ \\
 & $C_{31} = \frac{1}{ (\alpha_3 - \alpha_1)(\alpha_3 - \alpha_2)(\alpha_3 -
 \alpha_4)}$ \\
 & $C_{41} = \frac{1}{ (\alpha_4 - \alpha_1)(\alpha_4 - \alpha_2)(\alpha_4 -
 \alpha_3)}$ \\
 \hline

 $\frac{1}{(\mu - \alpha_1)(\mu - \alpha_2)(\mu - \alpha_3)^2}$
 & $C_{11} = \frac{1}{(\alpha_1 - \alpha_2)(\alpha_1 -
 \alpha_3)^2}$,
 \ $C_{21} = \frac{1}{(\alpha_2 - \alpha_1)(\alpha_2 - \alpha_3)^2}$ \\
 & $C_{31} = \frac{\alpha_1 + \alpha_2 - 2 \alpha_3}{(\alpha_3 - \alpha_1)^2(\alpha_3 -
 \alpha_2)^2}$,
 \ $C_{32} = \frac{1}{(\alpha_3 - \alpha_1)(\alpha_3 - \alpha_2)}$ \\
 \hline

 $\frac{1}{(\mu - \alpha_1)(\mu - \alpha_2)^3}$
 & $C_{11} =  \frac{1}{(\alpha_1 - \alpha_2)^3}$,
 \ $C_{21} =  - C_{11}$ \\
 & $C_{22} =  \frac{-1}{(\alpha_1 - \alpha_2)^2}$,
 \ $C_{23} =  \frac{1}{(\alpha_2 - \alpha_1)}$ \\
 \hline

 $\frac{1}{(\mu - \alpha_1)^2(\mu - \alpha_2)^2}$ &
 $C_{11} = - \frac{2}{(\alpha_1 - \alpha_2)^3}$,\
 $C_{12} = \frac{1}{(\alpha_1 - \alpha_2)^2}$,\
 \\
 & $C_{21} = - C_{11}$,\
 $C_{22} = - C_{12}$
 \\ \hline

\end{tabular}
\caption{Coefficients for terms potentially appearing in the expansion 
of the integral for the powder average of the LFC.} \label{expansion_coeff_1}
\end{table}

\begin{table}[t]
\begin{tabular}{|c|c|}
 \hline
 Form of Expansion Term & Expansion Coefficients
 \\ \hline \hline

 $\frac{1}{
     (\mu - \alpha_1)^2(\mu - \alpha_2)^2(\mu - \alpha_3)^2(\mu -
     \alpha_4)^2 }$
 & $C_{11} =  \frac{ -2 \left\{
    (\alpha_1 - \alpha_3)(\alpha_1 - \alpha_4) + (\alpha_1 -
    \alpha_2)(\alpha_1 - \alpha_4) + (\alpha_1 -
    \alpha_2)(\alpha_1 - \alpha_3) \right\}
    }{(\alpha_1 - \alpha_2)^3(\alpha_1 - \alpha_3)^3(\alpha_1 - \alpha_4)^3}$ \\
 & $C_{12} = \frac{1}{(\alpha_1 - \alpha_2)^2(\alpha_1 - \alpha_3)^2(\alpha_1 - \alpha_4)^2}$ \\
 & $C_{21} = \frac{ -2 \left\{
    (\alpha_2 - \alpha_3)(\alpha_2 - \alpha_4) + (\alpha_2 -
    \alpha_1)(\alpha_2 - \alpha_4) + (\alpha_2 -
    \alpha_1)(\alpha_2 - \alpha_3) \right\}
    }{(\alpha_2 - \alpha_1)^3(\alpha_2 - \alpha_3)^3(\alpha_2 - \alpha_4)^3}$ \\
 & $C_{22} = \frac{1}{(\alpha_2 - \alpha_1)^2(\alpha_2 - \alpha_3)^2(\alpha_2 - \alpha_4)^2}$ \\
 & $C_{31} = \frac{ -2 \left\{
    (\alpha_3 - \alpha_2)(\alpha_3 - \alpha_4) + (\alpha_3 -
    \alpha_1)(\alpha_3 - \alpha_4) + (\alpha_3 -
    \alpha_1)(\alpha_3 - \alpha_2) \right\}
    }{(\alpha_3 - \alpha_1)^3(\alpha_3 - \alpha_2)^3(\alpha_3 - \alpha_4)^3}$ \\
 & $C_{32} = \frac{1}{(\alpha_3 - \alpha_1)^2(\alpha_3 - \alpha_2)^2(\alpha_3 - \alpha_4)^2}$ \\
 & $C_{41} =  \frac{ -2 \left\{
    (\alpha_4 - \alpha_2)(\alpha_4 - \alpha_3) + (\alpha_4 -
    \alpha_1)(\alpha_4 - \alpha_3) + (\alpha_4 -
    \alpha_1)(\alpha_4 - \alpha_2) \right\}
    }{(\alpha_4 - \alpha_1)^3(\alpha_4 - \alpha_2)^3(\alpha_4 - \alpha_3)^3}$ \\
 & $C_{42} = \frac{1}{(\alpha_4 - \alpha_1)^2(\alpha_4 - \alpha_2)^2(\alpha_4 - \alpha_3)^2}$ \\
 \hline

 $\frac{1}{(\mu - \alpha_1)^2(\mu - \alpha_2)^2(\mu - \alpha_3)^4}$
 & $C_{11} = \frac{-2 \left\{ 3\alpha_1 - 2\alpha_2 - \alpha_3 \right\} }{(\alpha_1 - \alpha_2)^3(\alpha_1 - \alpha_3)^5} $,
 \ $C_{12} = \frac{1}{(\alpha_1 - \alpha_2)^2(\alpha_1 - \alpha_3)^4}$ \\
 & $C_{21} = \frac{-2 \left\{ 3\alpha_2 - 2\alpha_1 - \alpha_3 \right\} }{(\alpha_2 - \alpha_1)^3(\alpha_2 -
 \alpha_3)^5}$,
 \ $C_{22} = \frac{1}{(\alpha_2 - \alpha_1)^2(\alpha_2 - \alpha_3)^4}$ \\
 & $C_{31} = - C_{11} - C_{21} $ \\
 & $C_{32} = \frac{3(\alpha_3 - \alpha_1)^2 + 4(\alpha_3 - \alpha_1)(\alpha_3 - \alpha_2) + 3(\alpha_3 - \alpha_2)^2}
             {(\alpha_3 - \alpha_1)^4(\alpha_3 - \alpha_2)^4} $ \\
 & $C_{33} = \frac{-2 \left\{ \alpha_3 - \alpha_1 - \alpha_2 \right\} }{(\alpha_3 - \alpha_1)^3(\alpha_3 -
 \alpha_2)^3}$,
 \ $C_{34} = \frac{1}{(\alpha_3 - \alpha_1)^2(\alpha_3 - \alpha_2)^2}$ \\
 \hline

 $\frac{1}{(\mu - \alpha_1)^2(\mu - \alpha_2)^6}$
 & $C_{11} = \frac{-6}{(\alpha_1 - \alpha_2)^7}$,
 \ $C_{12} = \frac{1}{(\alpha_1 - \alpha_2)^6}$ \\
 & $C_{21} = -C_{11}$,
 \ $C_{22} = 5 C_{12}$,
 \ $C_{23} = \frac{-4}{(\alpha_2 - \alpha_1)^5}$ \\
 & $C_{24} = \frac{3}{(\alpha_2 - \alpha_1)^4}$,
 \ $C_{25} = \frac{-2}{(\alpha_2 - \alpha_1)^3}$,
 \ $C_{26} = \frac{1}{(\alpha_2 - \alpha_1)^2}$ \\
 \hline

 $\frac{1}{(\mu - \alpha_1)^4(\mu - \alpha_2)^4}$
 & $C_{11} = \frac{-20}{(\alpha_1 - \alpha_2)^7}$,
 \ $C_{12} = \frac{10}{(\alpha_1 - \alpha_2)^6}$,
 \ $C_{13} = \frac{-4}{(\alpha_1 - \alpha_2)^5}$,
 \ $C_{14} = \frac{1}{(\alpha_1 - \alpha_2)^4}$ \\
 & $C_{21} = -C_{11}$,
 \ $C_{22} = C_{12} $,
 \ $C_{23} = -C_{13}$,
 \ $C_{24} = C_{14}$ \\
 \hline
\end{tabular}
\caption{Coefficients for terms potentially appearing in the expansion 
of the integral for the powder average of the LFC$^2$.}
\label{expansion_coeff_2}
\end{table}


\begin{references}
\bibitem{aperature}
With such a fixed horizontal aperature, the azimuthal acceptance angle decreases as $1/q$, 
in principle distorting the line shape for even an extremely narrow 
aperature.  However, efforts to account for this distortion in the analysis of 
the smectic lineshapes showed that it had negligable effect.
\bibitem{quasi}
In the smectic phase, the correlations do not have true long-range 
order but exhibit algebraic decay due to the Landau-Peierls instability.  
We do not think this distinction should influence the 
form of the correlations in the presence of quenched 
disorder.
\bibitem{8s5fits} 
As for the fits made to 8CB + aerosil data, $\xi_{\bot}$ and $c$ were 
taken to have the same functional dependence on $\xi_{\|}$, 
$\xi_{\bot}(\xi_{\|})$ and $c(\xi_{\|})$, as in pure $\overline{8}$S5.
\end{references}
\end{document}